\documentstyle[twocolumn,aps,prb,epsfig,floats]{revtex}

\begin{document}
\twocolumn[\hsize\textwidth\columnwidth\hsize\csname@twocolumnfalse\endcsname

\title{Discreteness and entropic fluctuations in GREM-like systems}

\author{M. Sasaki$^1$ and O.~C.~Martin$^{1,2}$}

\address{
 $^{1}$ Laboratoire de Physique Th\'eorique et Mod\`eles Statistiques,
b\^at. 100, Universit\'e Paris-Sud, F--91405 Orsay, France.\\
 $^{2}$ Service de Physique de l'\'Etat Condens\'e\\
Orme des Merisiers --- CEA Saclay, 91191 Gif sur Yvette Cedex, France.\\
}

\date{\today}
\maketitle

\begin{abstract}
Within generalized random energy models,
we study the effects of energy discreteness
and of entropy extensivity in the low temperature phase. At zero temperature, 
discreteness of the energy
induces replica symmetry breaking, in contrast
to the continuous case where the ground state is unique.
However, when the ground state energy has an extensive entropy,
the distribution of overlaps $P(q)$ instead tends towards a single
delta function in the large volume limit. Considering
now the whole frozen phase, we find that
$P(q)$ varies continuously with temperature, and that
state-to-state fluctuations of entropy wash out
the differences between the discrete and continuous energy
models.
\end{abstract}

\pacs{PACS Numbers~: 75.10.Nr, 64.60.Cn, 64.70.Pf, 11.17.+y}

\twocolumn]\narrowtext

\section{Introduction}
\label{sect_intro}

Strongly disordered systems such as spin glasses~\cite{Young98}
have many metastable states in their low temperature
frozen phase, rendering them very sensitive to perturbations.
An extreme case of sensitivity arises in
systems having replica symmetry breaking~\cite{MezardParisi87b} (RSB): there
the excess free-energy
of the lowest excited states is $O(1)$ and so $P(q)$, the distribution
of overlaps $q$, is non-trivial. Given such sensitivity,
what properties are robust to changing the details of the 
microscopic Hamiltonian, and furthermore can
the presence of RSB itself depend on microscopic details?
Our purpose is to investigate this point in the context
of tractable models of spin glasses which are of the
mean field type. The main motivation for this is the 
question of RSB at $T=0$ in the $\pm J$ Ising spin
glass. It has been argued by Krzakala and 
Martin~\cite{KrzakalaMartin01} that in physical systems
with highly degenerate ground states there should be no replica
symmetry breaking at $T=0$ even if there is RSB at $T>0$.
Numerical investigations of this issue in the three-dimensional
Ising spin glass have led to conflicting claims, either
validating~\cite{PalassiniYoung01,HartmannRicci-Tersenghi02}
this picture or on the contrary~\cite{HedHartmann01}
suggesting that there
is RSB amongst the ground states of that model.
To understand better this question, Drossel and 
Moore~\cite{DrosselMoore01} have investigated overlaps in the $\pm J$ model
on the Migdal-Kadanoff
lattice at $T=0$ and $T>0$, concluding that 
$P(q=0)$ behaves differently with lattice size in the two
cases; however that model does not have RSB at any
temperature. In this work we provide a study
of this question in a solvable
model having RSB in its low temperature phase. 

This paper begins with the random energy model~\cite{Derrida81}
of Derrida and we consider the effects of
having discrete energies. Because the ground state can be
degenerate, $P(q)$ at $T=0$ has a strictly positive
weight at $q=0$ while in the continuous case $P(q=0)$ goes
to zero linearly with temperature. Then we consider 
generalized random energy models~\cite{Derrida85} and in particular
a discrete version with 
an infinite number of layers that can
be compared to the continuous version
studied by Derrida and Spohn~\cite{DerridaSpohn88}. 
In all these cases, the discreteness gives rise to RSB at $T=0$.
One of the unphysical features
of such models is their zero entropy density 
at low enough temperature.
Since we expect entropy fluctuations to be important for the
question of replica symmetry breaking, we extend this
last model so that states have random entropies.
For the models considered with an infinite number
of layers, we compute $P(q)$, the disorder averaged 
probability distribution of overlaps.
We then see the effects of
energy discreteness and of entropy fluctuations, in particular
as $T\to 0$. Finally, we consider how replica
symmetry is restored at $T=0$ in the thermodynamic limit
through entropic fluctuations, even though 
the energies are discrete.
 
\section{Random Energy Model} 
\label{sec:rem}
The REM~\cite{Derrida81}
is a simple model of spin glasses; its
partition function is
\begin{equation}
Z = \sum_{i=1}^{2^N} \exp \bigl[ -\frac{E_i}{T} \bigr] 
\end{equation}
where $N$ is identified with the number of 
spins of the system and the units are chosen so
that the Boltzmann constant is $1$. The energies
$E_i$ are independent identically distributed 
random variables of law $P_N(E)$. To 
make contact with physical systems,
$P_N(E)$ is set to coincide with the distribution of energies 
of spin configurations
in the $d$-dimensional Edwards-Anderson~\cite{EdwardsAnderson75} (EA)
spin glass. Such a system with $N$ spins
has $2^N$ energy levels but these energies are
strongly correlated; the REM is obtained if one
neglects these correlations.
If the couplings $J_{ij}$ of the EA model are Gaussian, 
then $P_N(E)$ is also Gaussian. If instead these couplings are
binary, $P(J_{ij})=\left[ \delta(J_{ij}+1) + \delta(J_{ij}-1) \right] /2$,
we are lead to a ``discrete'' REM where
$E$ is an integer random variable with distribution
$P_N(E=-dN+2k)=2^{-dN} {\rm B}(dN,k)$. In this expression, B is the 
usual binomial coefficient giving the number of ways
to choose $k$ elements out of $dN$.
This distribution is roughly Gaussian at large $N$.
The differences between the discrete and continuous models becomes
most apparent for the extreme energies and thus in the
low temperature phase where the lowest levels
dominate the partition function.

Let us begin with the properties at $T=0$. 
The ground state energy $E_0$ grows linearly with $N$ and
its variance is $O(1)$ in the large $N$ limit~\cite{Derrida81}.
The ground state is unique in the Gaussian case whereas
it has a strictly positive probability of being {\it degenerate}
in the discrete case.
This has important consequences.
Consider the overlap distribution $P_J(q)$
for a given disorder instance; when the ground state is
non-degenerate, $P_J$ will be a delta function at $q=1$,
while when it is degenerate, $P_J$ will have
two delta function peaks, one at $q=0$ and one at $q=1$. (By convention,
any two distinct levels are taken to have
zero overlap.) Since the degeneracy 
does not disappear at large $N$ as we show below, 
$P(q)$, the disorder average
of $P_J$, tends towards two delta function peaks, each of
weight $O(1)$ as $N \to \infty$. Thus at $T=0$ there is RSB
in the discrete REM but none in the Gaussian REM.

Now we shall be more quantitative and determine the
complete behavior of $P_J(q)$ at zero temperature. Indeed,
if the ground state is $g$-fold degenerate for a particular
disorder instance, one has 
\begin{equation}
P_J(q) = (1-g^{-1}) \delta (q) + g^{-1} \delta(q-1).
\label{eqn:PJQREM}
\end{equation}
The problem then reduces to computing the statistics of the integer $g$.
This can be done analytically as follows. 
Denote by $Q_N^{\rm gs}(n,g)$
the probability that the ground state energy 
is $-dN+2n$ and its degeneracy $g$.
In the discrete REM, we have
\begin{equation}
Q_N^{\rm gs}(n,g)={\rm B}(2^N,g)~p_N(n)^g~\left[\sum_{k=n+1}^{dN} p_N(k)\right]^{2^N-g},
\label{eqn:Pminnm}
\end{equation}
where we have introduced $p_N(n)=P_N(E=-dN+2n)$ to simplify the 
notation. Now we are interested in 
calculating the probability of the degeneracy 
\begin{equation}
Q_N^{\rm deg}(g)\equiv \sum_{n=0}^{dN} Q_N^{\rm gs}(n,g) .
\label{eqn:defQdeg}
\end{equation}
From that we shall calculate 
\begin{equation}
\langle g^{-1} \rangle \equiv \sum_{g=1}^{2^N} Q_N^{\rm deg}(g) g^{-1}
\end{equation}
which according to eq.~(\ref{eqn:PJQREM}) 
gives us the disorder averaged $P(q)$ of the model. 

Now let us calculate $Q_N^{\rm gs}(n,g)$ for $a(n)\equiv\frac{n}{dN}<0.5 $. 
We are not interested in the case $a(n)\ge 0.5$ because the probability that
the ground state energy is positive is negligible (recall that
$E$ and $n$ are related by $E=-dN+n$). 
Since
\begin{equation}
p_N(n-\Delta n)=
p_N(n)\left(\frac{a(n)}{1-a(n)}\right)^{\Delta n}
\left\{1+{\cal O}\left(\frac{1}{dN}\right)\right\}
\label{eqn:EXPapprox}
\end{equation}
holds for any finite integer $\Delta n$, the last factor in 
eq.~(\ref{eqn:Pminnm}) for $a(n)<0.5$ can be estimated as 
\begin{eqnarray}
&&\left[\sum_{k=n+1}^{dN} p_N(k)\right]^{2^N-g}\nonumber \\
&&=\left[1-\sum_{k=0}^{n} p_N(k)\right]^{2^N-g}\nonumber \\
&&\approx \left[1-\frac{[1-a(n)]p_N(n)}{1-2a(n)}\right]^{2^N-g} \nonumber \\
&&\approx \exp\left\{-\frac{[1-a(n)][2^N-g]p_N(n)}{1-2a(n)}\right\}\hspace{5mm}(a(n)<0.5).\end{eqnarray}
By substituting this equation into eq.~(\ref{eqn:Pminnm}), we find
\begin{equation}
Q_N^{\rm gs}(n,g) \approx \frac{1}{g!}\bigl[p_N(n)2^N\bigr]^g 
\exp\left\{-\frac{[1-a(n)]2^Np_N(n)}{1-2a(n)}\right\},
\label{eqn:Pmin2}
\end{equation}
provided that $a(n)<0.5$ and $g$ is not of order $\exp(N)$. 

Now let us denote by $n^*$ the integer for which 
the value of $p_N(n)2^N$ is the 
closest to $1$. Because eq.~(\ref{eqn:EXPapprox}) is also valid 
for $n=n^*$, $p_N(n^*+\Delta n)2^N$ for finite $\Delta n$ is expressed as
\begin{equation}
p_N(n^*+\Delta n)2^N=C^* \exp(\alpha^* \Delta n)
\left\{1+{\cal O}\left(\frac{1}{dN}\right)\right\} ,
\label{eqn:transient}
\end{equation}
where 
\begin{equation}
\alpha^*=\log\left(\frac{1-a^*}{a^*}\right),
\end{equation}
$a^*=a(n^*)$, and $C^*\equiv p_N(n^*)2^N$. Now the condition 
$p_N(n^*)2^N\approx 1$ determines $n^*$ and leads
to the following equation determining $a^*$ 
in the large $N$ limit:
\begin{equation}
(1-d)\log(2)-d\log(1-a^*)
+a^*\log\left(\frac{1-a^*}{a^*}\right)=0.
\end{equation}
The substitution of eq.~(\ref{eqn:transient}) into eq.~(\ref{eqn:Pmin2}) 
leads us to
\begin{eqnarray}
Q_N^{\rm gs}(n^*+\Delta n,g)& \approx &\frac{1}{g!}  (C^*)^g \exp(g\alpha^* \Delta n)
\nonumber \\
&&\times\exp\left\{-\frac{[1-a^*]C^*\exp(\alpha^* \Delta n)}{1-2a^*}\right\}
\nonumber \\
\end{eqnarray}
and these two quantities become equal as $N \to \infty$. 

To obtain a closed form expression for $\langle g^{-1} \rangle$, 
we now approximate the summation in eq.~(\ref{eqn:defQdeg}) by an integral 
and find in the large $N$ limit
\begin{eqnarray}
Q_N^{\rm deg}(g)&\approx& \int_{-\infty}^{\infty} {\rm d}x 
\frac{1}{g!}  (C^*)^g \exp(g\alpha^* x) \nonumber \\
&&\times\exp\left\{-\frac{[1-a^*]C^*\exp(\alpha^* x)}{1-2a^*}\right\}\nonumber \\
&\approx& \frac{1}{g\alpha^*}\left(\frac{1-2a^*}{1-a^*}\right)^g.
\label{eqn:finalQ_Ndeg}
\end{eqnarray}
This formula tells us that the probability that the ground state is degenerate 
is non-zero even in the limit $N\rightarrow \infty$
and that the distribution of $g$ is well behaved, i.e., 
all moments are finite. From eq.~(\ref{eqn:finalQ_Ndeg}), we finally obtain
\begin{equation}
\langle g^{-1} \rangle =\frac{1}{\alpha^*} \int_{0}^{\frac{1-2a^*}{1-a^*}}
\frac{{\rm d} x}{x}\log\left(\frac{1}{1-x}\right),
\end{equation}
where we have used 
\begin{equation}
\sum_{g=1}^{\infty} \frac{x^g}{g^2}=\int_0^x 
\frac{{\rm d} t}{t}\log\left(\frac{1}{1-t}\right)\hspace{5mm}(|x|\le 1).
\end{equation}
If we apply these expressions to our model when $d=3$, we
find $a^*=0.317$ and $\langle g^{-1} \rangle =0.824$; the peak
in the overlap distribution is thus much lower
at $q=0$ than at $q=1$. This finishes our analysis at $T=0$.

Consider now what happens at $T>0$. The detailed nature of $P_N(E)$
will remain important as long as the partition function
is dominated by a finite number of energy levels.
This happens throughout the whole low temperature
phase ($T < T_{\rm c}$) where
the free-energy and $P_J(q)$ are not self-averaging.
Consider in particular the $T$ 
dependence of $P(q)$ at low temperature. In the Gaussian case there
is a non-zero probability density to have a zero energy gap;
this leads to a $P(q,T)$ that is linear in $T$ as $T\to 0$.
On the contrary, in the discrete case we have
\begin{equation}
\label{eq_gap}
P(q=0,T) = P(q=0,T=0) + O(e^{-\frac{\Delta E}{T}}), 
\end{equation}
where $\Delta E=2$ is the energy gap of this model.
Indeed this holds for each disorder instance and so also holds 
for the disorder average.

\section{Generalized Random Energy Model}
\label{sec:grem}
In the GREM~\cite{Derrida85}
one considers $2^N$ states
that are organized in a tree-like fashion, allowing for correlated
energies.
At the first layer of the tree, there are $2^{N_1}$ branches; then each
such branch gives rise to $2^{N_2}$ branches in the second layer,
etc... The nodes of the last layer are identified with states
(or spin configurations), and there are $2^N$ of them where
$N=N_1 + N_2 \dots + N_L$. To each branch one associates
a random energy. Finally, the energy of a
state (leaf of the tree) is given by the sum of the energies
of the $L$ branches connecting it to the root of the tree (residing
at level $0$).

Just as for the REM, the low temperature phase of the GREM
is sensitive to the detailed distribution of the
energies. In the context of our study, we see that the
properties found for the REM extend to the
GREM as follows. At $T=0$, we need consider only the
ground states. At each level, there is a strictly positive probability to
have degenerate lowest energies for the 
model having discrete energies. One thus has a strictly
positive probability to find any of the possible overlaps
when considering ground states only. This is to be contrasted
with the Gaussian model for which $P(q) = \delta(q-1)$ when $T=0$.
Similarly, at $T>0$, $P(q)$ will
be quite sensitive to the nature of the energies
as long as we stay within the spin glass phase. At very low
temperatures, the weights of its peaks other than at $q=1$
will be linear in $T$ in the Gaussian case while they will
have an exponentially small temperature dependence in the
discrete case (cf. eq.~\ref{eq_gap}). Finally,
as one approaches the highest of the critical
temperatures, many branches at each level contribute to the partition
function and so from there on the detailed distribution
of energies becomes irrelevant.

\section{Infinite-level GREM}
\label{sec:tree}
\begin{figure}
\begin{center}
\epsfig{file=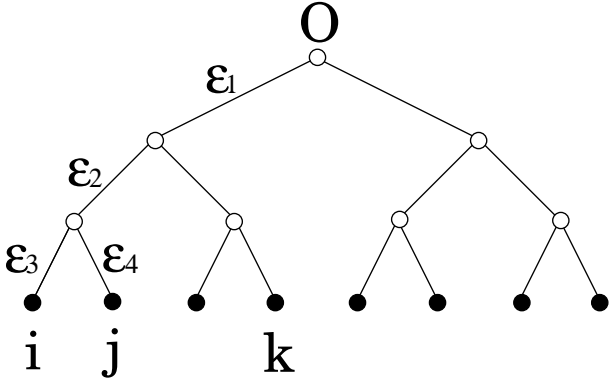,width=0.45\textwidth}
\end{center}
\caption{Construction of the infinite-level GREM with a branching factor $K=2$. 
}
\label{fig_tree}
\end{figure}

In a discrete GREM with $L$-layers, the ground state may be
degenerate for any finite $L$, but what happens when 
$L \to \infty$? To study that limit,
we set $N_i= k$, $k$ being a fixed integer (say $1$ or $2$). 
The model is schematically shown in figure~\ref{fig_tree}. 
There is then a fixed branching factor $K=2^k$ at each layer, each
node generating $K$ branches. A random energy $\epsilon$ is associated 
with each branch of the tree. The $\epsilon$ variables are independent and 
drawn from the same distribution $\rho_E(\epsilon)$. 
The energy $E(i)$ of a state $i$ is given 
by summing up the $\epsilon$'s of the branches which lie 
along the path connecting it to the tree's root $O$, e.g., 
in figure~\ref{fig_tree}, $E(i)=\epsilon_1+\epsilon_2+\epsilon_3$ 
and $E(j)=\epsilon_1+\epsilon_2+\epsilon_4$. 
The distance $d_{ij}$ of two states $i$ and $j$ is $d$ 
if their first common ancestor arises on the $d$-th layer counted from below, 
e.g., in figure~\ref{fig_tree}, $d_{ii}=0$, $d_{ij}=1$ and 
$d_{ik}=d_{jk}=2$. 
The overlap $q_{ij}$ is related to $d_{ij}$ by 
\begin{equation}
q_{ij}=1-d_{ij}/L. 
\label{eqn:overlap}
\end{equation}

This model has been studied in depth by Derrida and Spohn~\cite{DerridaSpohn88} 
(see also \cite{MajumdarKrapivsky00,DeanMajumdar01});
it can be viewed either as an infinite level
GREM or as a directed polymer on
a disordered Cayley tree. When the energy of each branch is taken from 
continuous distribution, the model's thermodynamics is extremely
close to that of the REM. There is a critical temperature below
which a finite number of states dominate the partition function,
and this low temperature phase exhibits one-step RSB, while the
distribution $P(q)$ tends towards a delta function at $q=1$ as $T \to 0$.


Since we are motivated by the question of RSB in the
$3$-dimensional $\pm J$ EA model, we shall investigate
the behavior of $P(q)$ for the Cayley tree model
when the energies on each branch are discrete. We shall 
take $\rho_E(\epsilon)$ to consist of a finite number of delta functions. 
To investigate the behavior of our model, we 
shall derive recursion formulae for the probability 
$Y(L,d)$ to find two states at a distance less or equal to $d$ by 
using the same techniques as developed
by Derrida and Spohn~\cite{DerridaSpohn88}. 
Since $q$ and $d$ are related by eq.~(\ref{eqn:overlap}), 
we can calculate $P(q)$ from $Y(L,d)$. 
By definition, the probability $Y(L,d)$ is given by
\begin{eqnarray}
Y(L,d)&\equiv& \overline{\frac{1}{Z(L)^2}
\sum_{ij/d_{ij}\le d}\exp[-X(i)-X(j)]}.
\label{eq_Y}
\end{eqnarray}
In this expression, $\overline{\cdots}$ represents the disorder average, 
$X(i)\equiv E(i)/T$, $Z$ is the partition function of the system, and
$L$ is the number of layers. A standard integral representation
for $Z^{-2}$ leads to
\begin{equation}
Y(L,d)=\int_{-\infty}^{\infty}{\rm d}u F(L,d;u),
\label{eqn:relationYF}
\end{equation}
where
\begin{eqnarray}
&&F(L,d;u)\nonumber\\
&&\equiv\overline{\exp[-2u-{\rm e}^{-u}Z(L)]
\sum_{ij/d_{ij}\le d} {\rm e}^{-X(i) -X(j)}}.
\label{eq_Y_int}
\end{eqnarray}
Collecting terms in the sum that belong to the same sub-tree of height $d$, 
we have
\begin{equation}
\sum_{ij/d_{ij}\le d} {\rm e}^{-X(i) -X(j)}=\sum_{B_d}\exp[-2X(B_d)] z(B_d)^2 .
\label{eqn:expression2}
\end{equation}
In this expression, $B_d$ is a general branch point 
in the $d$-th layer counted from below, 
$z(B)$ is the partition function of the sub-tree rooted at 
a branch point $B$, $X(B)\equiv E(B)/T$, and 
$E(B)$ is the energy of a branch point $B$ given by 
summing up the $\epsilon$'s of the branches which lie 
along the path connecting $B$ and $O$. 
Substitution of eq.~(\ref{eqn:expression2}) 
into eq.~(\ref{eq_Y_int}) gives
\begin{eqnarray}
&&F(L,d;u)\nonumber \\
&&=\overline{\exp[-2u-{\rm e}^{-u}Z(L)]
\sum_{B_d}\exp[-2X(B_d)] z(B_d)^2}.
\label{eqn:iniEQforF}
\end{eqnarray}
From this equation, we find
\begin{equation}
F(d,d;u)=H_2(d;u),
\label{eqn:initialF}
\end{equation}
where
\begin{equation}
H_n(d;u)\equiv \overline{\{{\rm e}^{-u}z(B_d)\}^n\exp[-{\rm e}^{-u}z(B_d)]}.
\label{eqn:defHn}
\end{equation}

We can calculate $H_n(d;u)$ 
(including $H_2(d;u)$ appearing in eq.~(\ref{eqn:initialF})) 
by the following recursion formulae. Now let us start from the 
simplest case, i.e., $n=0$. By definition, 
\begin{equation}
H_0(0;u)=\exp[-e^{-u}].
\label{eqn:Hrecursion1}
\end{equation}
Since the $z(B_{d+1})$ can be expressed as 
\begin{equation}
z(B_{d+1})=\sum_{B_d} {\rm e}^{-X(B_d)}z(B_d), 
\end{equation}
and $z(B_d)$ and $z(B_d')$ are independent if $B_d\ne B_d'$, 
we obtain the recursion formula
\begin{eqnarray}
H_0(d+1;u)&=& \prod_{B_d} \overline{\exp[-{\rm e}^{-u-X(B_d)}z(B_d)]}\nonumber \\
&=& {\tilde H}_0(d;u)^K,
\label{eqn:Hrecursion2}
\end{eqnarray}
where we have defined 
\begin{equation}
{\tilde g}(u)\equiv \int {\rm d}\epsilon 
\rho_E(\epsilon)g(u+\epsilon/T)
\label{eqn:deftilde}
\end{equation}
for a general function $g(u)$. 
The recursion formulae for $n\ne 0$ are derived 
by using eqs.~(\ref{eqn:Hrecursion1}) and (\ref{eqn:Hrecursion2}) as well as 
the relation 
\begin{equation}
H_n(d;u)=\frac{{\rm d}^n}{{\rm d} u^n} H_0(d;u).
\label{eqn:formulaH_n}
\end{equation}
For example, the recursion formula for $H_1(d;u)$ is
\begin{eqnarray}
H_1(d+1;u)&=&\frac{\rm d}{{\rm d} u}{\tilde H_0}(d;u)^K\nonumber \\
&=&  K{\tilde H_1}(d;u) {\tilde H_0}(d;u)^{K-1}.
\end{eqnarray}

Finally, let us derive recursion formulae for $F(L,d;u)$. 
From eq.~(\ref{eqn:iniEQforF}), we have 
\begin{eqnarray}
&&F(L+1,d;u)\nonumber \\
&&=\overline{\exp\biggl[-2u-{\rm e}^{-u}\sum_{B_L'}{\rm e}^{-X(B_L')}z(B_L')
\biggr]}\nonumber \\
&&\quad\times\sum_{B_L}\sum_{B_d\in B_L} \overline{\exp[-2X(B_d)] z(B_d)^2}
\nonumber \\
&&=\sum_{B_L} \overline{\exp\left[-2u-2X(B_L)-{\rm e}^{-u-X(B_L)} z(B_L)\right]}
\nonumber \\
&&\quad\times\sum_{B_d\in B_L}\overline{{\rm e}^{-2[X(B_d)-X(B_L)]}z(B_d)^2}
\nonumber \\
&&\quad\times\prod_{B_L'\ne B_L} \overline{\exp\left[-{\rm e}^{-u-X(B_L')}
z(B_L')\right]}.
\end{eqnarray}
The strategy now is to perform the disorder average in two steps:
first we average over sub-trees rooted at $\{B_L\}$ when fixing $\{X(B_L)\}$; 
then we average over $\{X(B_L)\}$. By using eq.~(\ref{eqn:iniEQforF}) and 
eq.~(\ref{eqn:defHn}) with $n=0$, the disorder average in the first step is done as
\begin{eqnarray}
F(L+1,d;u)&=&\sum_{B_L} \overline{F(L,d;u+X(B_L))}\nonumber \\
&&\times\prod_{B_L'\ne B_L} \overline{H_0(L;u+X(B_L'))}.
\end{eqnarray}
Note that the only random variables in this equation are $\{ X(B_L)\}$. 
The disorder average in the next step finally leads us to
\begin{equation}
F(L+1,d;u)=K {\tilde F}(L,d;u) {\tilde H}_0(L;u)^{K-1}.
\label{eqn:finalresult}
\end{equation}

In summary, the disorder averaged distribution of distances
$Y(L,d)$ can be computed by the following procedures:
\begin{itemize}
\item[i)] Calculate $H_2(d;u)$ (=$F(d,d;u)$) by evaluating numerically 
the recursions which are derived by applying eq.~(\ref{eqn:formulaH_n}) 
to eqs.~(\ref{eqn:Hrecursion1}) and~(\ref{eqn:Hrecursion2}). 
\item[ii)] Calculate $F(L,d;u)$ by using the recursion eq.~(\ref{eqn:finalresult}). 
\item[iii)] Compute $Y(L,d)$ by estimating numerically the integral 
in eq.~(\ref{eqn:relationYF}). 
\end{itemize}
For large number of layers we see convergence to the large $L$ limit
after introducing the (continuous) overlap $q = 1-d/L$;
this gives us the disorder average $P(q)$
or equivalently $\overline{Y(q)} = \int_0^q \overline {P(q)} {\rm d}q$
for the infinite tree. 

\begin{figure}
\begin{center}
\epsfig{file=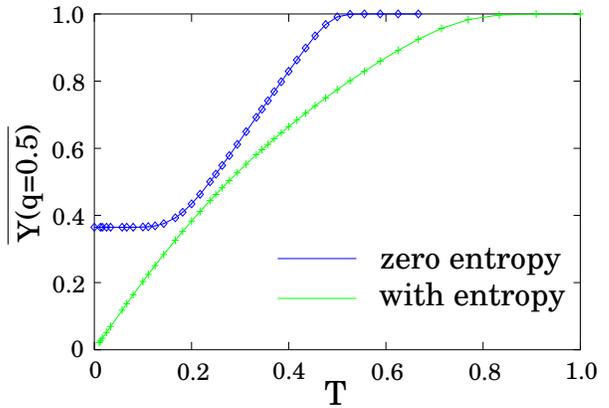,width=0.45\textwidth}
\end{center}
\caption{$\overline {Y(q=0.5)}= \int_{0}^{0.5} P(q) {\rm d}q$ as a function of
temperature, for the infinite level GREM (top curve) and for 
its random entropy generalization (bottom curve). (Plots are
for $K=2$ and $L=1000$.)}
\label{fig_YofT}
\end{figure}

Just as for the Gaussian case, the discrete model has one step
RSB below a critical temperature $T_{\rm c}$. 
However, the energies are discrete and
we find that the degeneracy of the ground state is maintained as
$L\to \infty$ and furthermore its mean
saturates to a finite value. Because of this,
there is RSB at $T=0$, not only for
$0<T<T_{\rm c}$. When executing our recursions, the distribution of overlaps
rapidly has very narrow peaks at $q=0$ and $q=1$ with
increasing number of layers, while
the probability of having overlaps between those two 
values goes to zero.
One can thus characterize the mean probability distribution
of overlaps by the weight of the delta function at
$q=0$. For that, we consider the quantity
$\overline {Y(q=0.5)}$ which gives the weight of
the overlaps near $q=0$.
In figure~\ref{fig_YofT} we
show the temperature dependence of that quantity 
(top curve). For these data, we used $K=2$ and 
\begin{equation}
\rho_E(\epsilon)=0.25\delta(\epsilon)+0.5\delta(\epsilon-1)
+0.25\delta(\epsilon-2).
\label{eqn:rhoE}
\end{equation}
Furthermore, we took $L=1000$ which is big enough to represent the
infinite tree limit at all $T$ except for $T$ close
to $T_{\rm c}$. We have checked that the curves near $T_{\rm c}$ 
converge as $L$ increases, and 
a cusp is created at $T_{\rm c}$ in the limit $L\rightarrow \infty$, 
just as the continuous case\cite{DerridaSpohn88}. 
But the point we want to make here concerns $T \ll T_{\rm c}$: 
the curve is very flat at low temperatures. This is as expected
from the analogy with the discrete REM (cf. eq.~\ref{eq_gap}),
and is due to the presence of a gap above the ground state
energy.

\section{A model with extensive entropy}
\label{sec:treeEntropy}
In all the models considered so far, a finite number of the lowest
energy states dominate the partition function at sufficiently
low temperatures. 
Thus the entropy is $O(1)$ rather than $O(N)$. Clearly, in any physical system,
the entropy remains extensive as long as $T>0$. What is
the possible importance of such extensivity? Krzakala
and Martin~\cite{KrzakalaMartin01} argued that extensive
entropies will give rise to diverging entropic fluctuations, and that
these fluctuations should prevent RSB at $T=0$.
To test this idea, one needs a model with extensive entropies
at low temperature. There are several ways to do this
in the context of GREM-like models; Cook
and Derrida~\cite{CookDerrida89} added small loops
to the tree, while Yoshino~\cite{Yoshino97} replaced the random
energies by random entropies. We keep the tree structure
but have both energies and entropies.

We focus again on the infinite tree with
a constant branching factor $K$. To each branch we assign
a random energy $\epsilon$ {\it and} a random entropy $\sigma$
from distributions $\rho_E(\epsilon)$ and $\rho_S(\sigma)$, respectively; 
then each leaf has an energy and entropy given by the sum of those terms
along the path connecting that leaf to the tree's root $O$.
In this model, both the energy and the entropy of the states become extensive.
Now a crucial point is that the model with and without entropies
can be mapped onto one-another when $T\ne 0$; indeed the two models
are {\it identical} provided the distributions satisfy 
\begin{equation}
\rho_E^*(\epsilon')=\int {\rm d}\epsilon {\rm d} \sigma 
\delta (\epsilon'-\epsilon+T\sigma)\rho_E(\epsilon)\rho_S(\sigma),
\end{equation}
where $\rho_E^*(\epsilon')$  is the distribution of the energy 
for the model without entropy. 
This condition insures that the distribution of the weight 
assigned to each branch ($\exp(-\epsilon/T+\sigma)$ 
and $\exp(-\epsilon'/T)$) is the same in the both cases. 
Therefore, we can utilize all of the recursions previously derived 
just by changing eq.~(\ref{eqn:deftilde}) into
\begin{equation}
{\tilde g}(u)\equiv \int {\rm d}\epsilon {\rm d}\sigma
\rho_E(\epsilon)\rho_S(\sigma)g(u+\epsilon/T-\sigma).
\end{equation}

We have investigated the distribution of overlaps in this
model with entropic fluctuations. We have chosen 
$\rho_E$ as before and set
\begin{equation}
\rho_S(\sigma) = 0.5\delta(\sigma) + 0.5\delta(\sigma-2).
\end{equation}
We have also chosen the same branching factor as before, i.e., $K=2$. 
The inclusion of these entropic
fluctuations changes the critical temperature,
but it also changes the qualitative behavior of RSB.
In figure~\ref{fig_YofT} we show the same quantity as before,
$\overline {Y(q=0.5)}$. We see that
the entropic fluctuations destroy RSB at $T=0$, and 
$P(0)$ goes to zero {\it linearly} 
with $T$ as $T\to 0$. One may think from the figure that 
there is no cusp at $T=T_{\rm c}$ in the model 
with entropy. However it can be shown from 
the mapping mentioned in the previous 
paragraph that the inclusion of entropy does {\it not} 
remove the cusp. To check this,
we have verified numerically that the 
cusp appears slowly as $L$ is increased and that 
the left derivatives of 
$\overline {Y(q=0.5)}$ at $T={T_{\rm c}}$
tend to a non-zero limit 
for both models as $L\rightarrow\infty$.

\begin{figure}
\begin{center}
\epsfig{file=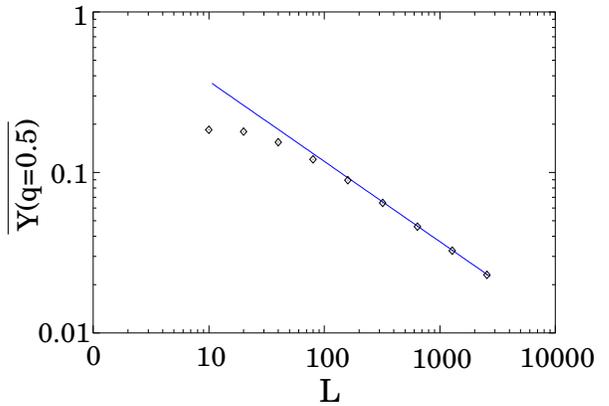,width=0.45\textwidth}
\end{center}
\caption{$\overline {Y(q=0.5)}$ at $T=0$ vs. $L$. 
The parameters for these data are $K=2$, $\rho_S(\sigma) = 0.5\delta(\sigma) 
+ 0.5\delta(\sigma-1)$ and eq.~(\ref{eqn:rhoE}) for $\rho_E$. 
A function proportional to $L^{-0.5}$ is drawn to guide the eye. 
}
\label{fig_PQT0}
\end{figure}

We have also investigated the finite size dependence of $P(q=0)$ at 
$T=0$. From figure~\ref{fig_PQT0}, we see that this quantity decreases 
as $L^{-0.5}$. This result can be understood from 
the following simple argument. Imagine a disorder instance
in which there are only two ground states 
whose overlap is close to $0$. The probability for an overlap
to be small for such a sample is
of order $1$ if the difference of the entropies of the two ground states 
is of order $1$. Recall that each ground
state has an extensive entropy; furthermore,
the entropy fluctuations of a ground state 
are necessarily of order $L^{0.5}$. Then
the probability that the two states have the {\it same} entropy 
is of order $L^{-0.5}$ in the large $L$ limit. From this we
conclude that at zero temperature, $P(q=0)$ decreases as $L^{-0.5}$
as $L \to \infty$.

It is worth pointing out that Krzakala and Martin~\cite{KrzakalaMartin01} 
suggested that in the $3-d$ EA $\pm J$ 
spin glass valley-to-valley fluctuations 
in the entropy should grow as $L^{d_{\rm s}/2}$ 
($d_{\rm s}$ is the fractal dimension of surfaces of large-scale 
low-energy excitations) and $P(0)$ should then
decrease as $L^{-d_{\rm s}/2}$. A
power law decay of $P(0)$ in that model 
has been found numerically~\cite{Hartmann00,PalassiniYoung01}. The 
contribution of our study is to show that such a phenomenon
does indeed occur beyond reasonable doubt in a particular model.

\section{Discussion}
\label{sec:discussion}

Let us compare the model with discrete energies and
extensive entropies to the one where
the disorder variables are continuous (be there entropy fluctuations
or not). In both cases, $P(q=0)$ goes
to zero linearly with $T$ as $T\to 0$ because the lowest
{\it free-energy} states are non-degenerate and there is no
gap. We may then summarize what we have found by saying that
state-to-state entropy fluctuations of order $\sqrt{L}$ 
effectively remove both degeneracies
and gaps. Of course at $T=0$, the degeneracy in the
energy is important, but our point is that the free-energies 
are still not degenerate:
the contribution of different ground
states to the partition
function are wildly different simply because their 
entropy differences diverge as $\sqrt L$. This is
the mechanism behind no RSB at zero temperature. 
If we consider now $T>0$, we note that
the energies of the states dominating
the partition function are far above the lowest energy;
the system selects the states with the minimum
{\it free-energies}, $F=E - T S$; even though both $E$ and $S$
are integers, generically $F$ is not and so
any fine structure of $E$ and $S$ is washed out. 

Our work has focused on the infinite tree model because it
is tractable, but we expect the conclusions 
to be quite general when there is one-step RSB at
$T>0$. An open question concerns of course the case
of systems having continuous RSB. One way to address
this is to generalize our model so that it exhibits
continuous RSB; this can be done just
as for the Derrida-Spohn model~\cite{DerridaSpohn88}
by making the distributions of energies
depend on the level of the tree.
With such a modification, the model having no entropy
but discrete energies will
have continuous RSB at $T=0$, whereas if random entropies are 
assigned to each branch the
entropic fluctuations will restore replica symmetry at $T=0$.

One may also ask what effect would $J_{ij}=\pm 1$ have on the
Sherrington-Kirkpatrick~\cite{SherringtonKirkpatrick75} 
(SK) model, i.e., would such
discreteness give rise to RSB at $T=0$, in contrast to what
happens in the Gaussian case? The crucial property that can make
RSB possible at $T=0$ is ground state degeneracy; the gap in such
a discrete SK model of $N$ spins is $1/\sqrt{N}$ if the $J_{ij}$s
are rescaled so that the model has a thermodynamic limit. 
The valley-to-valley energy differences being $O(1)$, 
the probability of an exact ground state degeneracy should 
be $O(1/\sqrt{N})$, so we do {\it not} expect RSB at $T=0$ here.
In fact, it seems unlikely that the discrete and continuous SK 
models have any differences in the thermodynamic limit because
the local fields on the spins effectively become continuous
in the large $N$ limit.

Finally, let us note that
the most controversial case of 
zero-temperature RSB arises in the $3-d$ EA 
$\pm J$ spin glass; its breaking of replica symmetry at $T>0$ is 
probably continuous, though even that is subject to
debate. It would thus be very useful to 
pursue this issue further to resolve the contradictory 
results~\cite{Hartmann00,PalassiniYoung01,HedHartmann01,HartmannRicci-Tersenghi02}
obtained so far. 

\section*{Acknowledgments ---}

We thank J.-P. Bouchaud, F. Krzakala and
M. M\'ezard for stimulating discussions.
M.S. acknowledges financial support from the MENRT.
The LPTMS is an Unit\'e de Recherche
de l'Universit\'e Paris~XI associ\'ee au CNRS.

\bibliographystyle{prsty}
\enlargethispage{30pt}

\bibliography{references}

\addcontentsline{toc}{chapter}{\protect\bibname}
\begin{thebibliography}{10}

\bibitem{Young98}
{\em Spin Glasses and Random Fields}, edited by A.~P. Young (World Scientific,
  Singapore, 1998).

\bibitem{MezardParisi87b}
M. M{\'e}zard, G. Parisi, and M.~A. Virasoro, {\em Spin-Glass Theory and
  Beyond}, Vol.~9 of {\em Lecture Notes in Physics} (World Scientific,
  Singapore, 1987).

\bibitem{KrzakalaMartin01}
F. Krzakala and O.~C. Martin, Europhys. Lett. {\bf 6},  749  (2001).

\bibitem{PalassiniYoung01}
M. Palassini and A.~P. Young, Phys. Rev. B {\bf 63},  140408(R)  (2001),
  cond-mat/0012164.

\bibitem{HartmannRicci-Tersenghi02}
A. Hartmann and F. Ricci-Tersenghi,   (2001), cond-mat/0108307.

\bibitem{HedHartmann01}
G. Hed, A. Hartmann, and E. Domany, Europhys. Lett. {\bf 55},  112  (2001),
  cond-mat/0012451.

\bibitem{DrosselMoore01}
B. Drossel and M. Moore, Eur. Phys. J. B {\bf 22},  589  (2001).

\bibitem{Derrida81}
B. Derrida, Phys. Rev. B {\bf 24},  2613  (1981).

\bibitem{Derrida85}
B. Derrida, J. Phys. Lett. (France) {\bf 46},  401  (1985).

\bibitem{DerridaSpohn88}
B. Derrida and H. Spohn, J. Stat. Phys. {\bf 51},  817  (1988).

\bibitem{EdwardsAnderson75}
S.~F. Edwards and P.~W. Anderson, J. Phys. F {\bf 5},  965  (1975).

\bibitem{MajumdarKrapivsky00}
S.~N. Majumdar and P.~L. Krapivsky, Phys. Rev. E {\bf 62},  7735  (2000).

\bibitem{DeanMajumdar01}
D. Dean and S. Majumdar, Phys. Rev. E {\bf 64},  046121  (2001).

\bibitem{CookDerrida89}
J. Cook and B. Derrida, Europhys. Lett {\bf 10 (3)},  195  (1989).

\bibitem{Yoshino97}
H. Yoshino, J. Phys. A. {\bf 30},  1143  (1997).

\bibitem{Hartmann00}
A. Hartmann, Eur. Phys. J. B {\bf 13},  539  (2000), cond-mat/0107308.

\bibitem{SherringtonKirkpatrick75}
D. Sherrington and S. Kirkpatrick, Phys. Rev. Lett. {\bf 35},  1792  (1975).

\end{thebibliography}

\end{document}